\documentclass[aps,onecolumn]{revtex4-1}
\usepackage{color}
\usepackage{soul}
\usepackage{dcolumn}
\usepackage{graphicx}
\usepackage{amsmath}
\usepackage{amssymb}
\usepackage{bm}

\begin{document}

%\preprint{}

%\title{Which features of the quantum Rabi model survive \\ in a multilevel circuit implementation with arbitrary coupling ?}

%\title{Circuit implementations of the quantum Rabi model with arbitrary strong coupling}

\title{Resilience of the quantum Rabi model in circuit QED }


\author{Vladimir E. Manucharyan}
\affiliation{Department of Physics, Joint Quantum Institute, and Center for Nanophysics and Advanced Materials,
University of Maryland, College Park, MD 20742, USA.}
\author{Alexandre Baksic}
\affiliation{Department of Physics, McGill University, 3600 rue University, Montreal, Qc H3A 2T8, Canada.}
\author{Cristiano Ciuti}
\affiliation{Laboratoire Mat\'eriaux et Ph\'enom\`enes Quantiques, Universit\'e Paris Diderot, CNRS UMR 7162, Sorbonne Paris Cit\'e, 10 rue Alice Domon et Leonie Duquet 75013 Paris, France.}

%\thanks{}


\date{\today}

\begin{abstract}
In circuit quantum electrodynamics (circuit QED), an artificial "circuit atom" can couple to a quantized microwave radiation much stronger than its real atomic counterpart. The celebrated quantum Rabi model describes the simplest interaction of a two-level system with a single-mode boson field. When the coupling is  large enough, the bare multilevel structure of a realistic circuit atom cannot be ignored even if the circuit is strongly anharmonic. We explored this situation theoretically for flux (fluxonium) and charge (Cooper pair box) type multi-level circuits tuned to their respective flux/charge degeneracy points. We identified which spectral features of the quantum Rabi model survive and which are renormalized for  large coupling. 
Despite significant renormalization of the low-energy spectrum in the fluxonium case, the key quantum Rabi feature -- nearly-degenerate vacuum consisting of an atomic state entangled with a multi-photon field -- appears in both types of circuits when the coupling is sufficiently large. 
Like in the quantum Rabi model, for very large couplings the entanglement spectrum is dominated by only two, nearly equal eigenvalues, in spite of the fact
that a large number of bare atomic states are actually involved in the atom-resonator ground state. We interpret the emergence of the two-fold degeneracy of the vacuum of both circuits as an environmental suppression of flux/charge tunneling due to their dressing by virtual low-/high-impedance photons in the resonator. For flux tunneling, the dressing is nothing else than the shunting of a Josephson atom with a large capacitance of the resonator. Suppression of charge tunneling is a manifestation of the dynamical Coulomb blockade of transport in tunnel junctions connected to resistive leads. 
\end{abstract}

%\keywords{}

\maketitle

\section{Introduction}

In the last decade, the field of circuit quantum electrodynamics (circuit QED) \cite{10.1038/nature02851,schoelkopf2008wiring} has emerged as a frontier of quantum information science thanks to the simultaneous combination of scalable fabrication, strong interactions, and high-coherence offered by superconducting circuits~\cite{devoret2013superconducting}.
% \st{vibrant field of research thanks to the possibility of creating highly-coherent artificial circuit atoms controllably coupled  to microwave resonators for quantum information applications and for the exploration of fundamental quantum phenomena.} 
In both circuit QED and cavity QED \cite{RevModPhys.73.565}, the resonant coupling between a single photon and a two-level atom is quantified by the so-called
vacuum Rabi frequency $g$, which is typically orders of magnitude smaller than the photon frequency $\omega_r$ and atom transition frequency $\omega_a$. 
Such a system is then well described by the Jaynes-Cummings Hamiltonian~\cite{jaynes1963comparison}, which is the rotating-wave  approximation of the celebrated quantum version of the Rabi model \cite{PhysRev.49.324,PhysRev.51.652}. However, in situations where  $g \sim \omega_r,\omega_a$, those systems cannot be accurately described by the Jaynes-Cummings model and one needs to consider the quantum Rabi model. The Hamiltonian corresponding to this model reads:
\begin{equation}
\label{Rabi}
H_{\rm Rabi} /\hbar = \omega_r a^{\dagger} a + \cfrac{\omega_{a}}{2} \sigma_z + g (a + a^{\dagger}) \sigma_x, 
\end{equation} \\
where $a^{\dagger}$ is a bosonic creation operator for a single electromagnetic mode with a quantum of energy $\hbar \omega_r$. The boson mode is linearly coupled to a two-level system whose algebra is described by the Pauli matrices $\sigma_x$, $\sigma_y$, $\sigma_z$.  In spite of its apparent simplicity, the exact analytical solution of the quantum Rabi model has been found only recently \cite{PhysRevLett.107.100401}.
% \st{Knowing that $\sigma_x = \sigma_+ + \sigma_-$, where $\sigma_+$ is the raising operator of the two-level system and $\sigma_-$ the lowering one, then it is clear that quantum Rabi Hamiltonian contains non-rotating wave coupling terms $g a \sigma_-$ and $g a^{\dagger} \sigma_+$. Dropping these two terms, the quantum Rabi Hamiltonian reduces to the Jaynes-Cummings model.}

An intriguing property of the quantum Rabi model (Eq.~\ref{Rabi}) in the ultrastrong coupling regime, i.e. when the condition $g\ll \omega_r$ is violated, is the emergence of a non-trivial ground state~\cite{PhysRevB.72.115303,PhysRevLett.104.023601,PhysRevLett.105.263603,PhysRevLett.104.023601,PhysRevA.81.042311,PhysRevLett.112.173601,PhysRevLett.111.243602}, which is not the ordinary vacuum.  In the standard scenario of cavity QED, while the excitations are entangled states of the atom and resonator, the ground state is the standard vacuum, namely the product of the zero photon Fock state and the atomic ground state.
Instead, in the quantum Rabi model for large enough coupling the ground state asymptotically becomes a Schr\"odinger cat state where the photon field is maximally entangled with the atom~\cite{hines2004entanglement,PhysRevLett.104.023601}.
The first excited state tends to an orthogonal entangled cat state. Moreover, the frequency gap between these two non-classical states decreases
exponentially with increasing coupling. A qubit consisting of these two entangled atom-resonator states is expected to enjoy protection with respect
to a class of decoherence channels and might have applications in quantum information \cite{PhysRevLett.107.190402}.

%\vlad{SHOULDN'T WE MOVE THIS TO CONCLUSIONS SECTION? INSTEAD, CAN YOU SAY WHAT ELSE IS INTERESTING ABOUT $g >\omega_r$?} It is worth pointing out that photon population in a ground state can not be released out of the resonator if the system Hamiltonian is time-independent \cite{PhysRevA.74.033811}. An efficient release of such ground state photons (quantum vacuum radiation) can occur only if the system parameters are changed in a non-adiabatic fashion \cite{PhysRevLett.98.103602,PhysRevA.80.053810,Gunter2009,Peropadre10}. Non-destructive measurement of the non-trivial ground state properties can be done for example using an ancillary qubit \cite{PhysRevLett.114.183601,FelicettiSciRep}.

In a recent experiment\cite{Langford2016}, an effective quantum Rabi dynamics  with large effective couplings has been realized by using digital quantum simulation techniques\cite{Digital} with time-dependent drives in a conventional circuit QED system with a relatively small vacuum Rabi coupling. In that driven configuration, one can realize states sharing distinctive properties of the quantum Rabi model eigenstates. Can one implement $g/\omega_r \gg 1$ in a system with a time-independent Hamiltonian? This is of course forbidden for real atoms due to the small value of the fine structure constant. Interestingly, artificial circuit atoms do not suffer from this limitation partly due to the possibility of direct interconnection of circuits by wires~\cite{ANDP:ANDP200710261}. Several groups utilized this approach and reported spectroscopic signatures of a regime where $g/\omega_r \sim 1$ with a superconducting flux qubit coupled to low-impedance resonant modes~\cite{PhysRevLett.105.237001,Niemczyk2010,yoshihara2016superconducting}.

Here we consider theoretically time-independent analog circuit implementations of the quantum Rabi model that allows to reach coupling  $g/\omega_r $ up to a value of the order of $10$.  When the coupling largely exceeds both the photon and bare two-level transition frequencies, one might wonder if the quantum Rabi model can adequately describe a circuit QED system. In particular, we address the two following problems. First, atom-photon interaction necessarily comes with "$A^2$"-like term (terms quadratic in the photon operators), which can be safely neglected  for the case of a single artificial atom and  $g/\omega_r \ll 1$. This procedure instead is not straightforward once this condition is violated, as it has been for example considered for the Dicke-like problem of $N \gg 1$ artificial, identical atoms coupled to the same resonator mode~\cite{nataf2010no,PhysRevLett.104.023601,viehmann2011superradiant,jaako2016ultrastrong,PhysRevLett.117.173601,PhysRevA.93.012120} (one of the inherent difficulties of the collective coupling of $N$ artificial atoms is the plain fact that an exact diagonalization of the circuit Hamiltonian including all relevant levels of the artificial atoms is not feasible for $N \gg 1$). Second, artificial circuit atoms do have more than two levels. Even at $g/\omega_r \ll 1$ one must take higher circuit levels into account, for instance, to obtain correct dispersive shifts at the second order in $g/\omega_r$. One can imagine that higher levels are even more important when $g/\omega_r \gg 1$. However, both questions remained largely unexplored in the previous studies investigating one circuit atom coupled to resonator mode~\cite{ANDP:ANDP200710261,PhysRevA.80.032109,peropadre2013nonequilibrium}. In this paper, we have identified which properties of the quantum Rabi model survive and which ones are modified when considering a multilevel circuit atom coupled  to a lumped-element single-mode resonator. 

The manuscript is organized as follows. In Sec. \ref{linear}, as a prelude  we consider the coupling between two linear $LC$-circuits and show that a regime where the coupling is larger than the transition frequency can be naturally achieved with conventional circuit elements by an appropriate interconnection choice. We describe this system both in the flux and charge gauge. We show that the corresponding quantized Hamiltonian is equivalent to the Hopfield Hamiltonian for semiconductor polaritons \cite{PhysRev.112.1555,PhysRevB.72.115303}.  In Sec. \ref{atom}, we introduce the Hamiltonian theory for a multilevel circuit atom  coupled to a single-mode lumped-element resonator. In particular, we separately consider the case of a fluxonium~\cite{Manucharyan113} artificial atom, which is based on the flux tunneling in and out of a superconducting loop and a Cooper pair box~\cite{nakamura1999coherent} artificial atom, which is based on charge tunneling in and out of a superconducting island. The two circuits are different by the topology of their circuit variables~\cite{koch2009charging} and together span the entire range of existing circuit atoms. Fluxonium results are qualitatively applicable to the flux qubit, while the Cooper pair box analysis covers the transmon qubit~\cite{koch2007charge} as well (indeed, in our theory we do not perform a two-level approximation, but consider all the levels necessary to get convergence of the observables). In both cases, we explore numerically (Sec. \ref{discussion}) the behavior of the transition spectrum, the vacuum degeneracy, and the atom-photon entanglement, drawing a comparison with the ideal two-level quantum Rabi model. We also provide a physical interpretation of the physical results in the limit $g/\omega_r \gg 1$, in terms of environmental suppression of flux and charge tunneling~\cite{clarke1988quantum, devoret1990effect}. Conclusions and perspectives are summarized in Sec. \ref{conclusions}.

\section{Prelude: how strong can the coupling between two linear circuits be?}
\label{linear}
\begin{figure*}[t!]
	\centering
	\includegraphics[width=\linewidth]{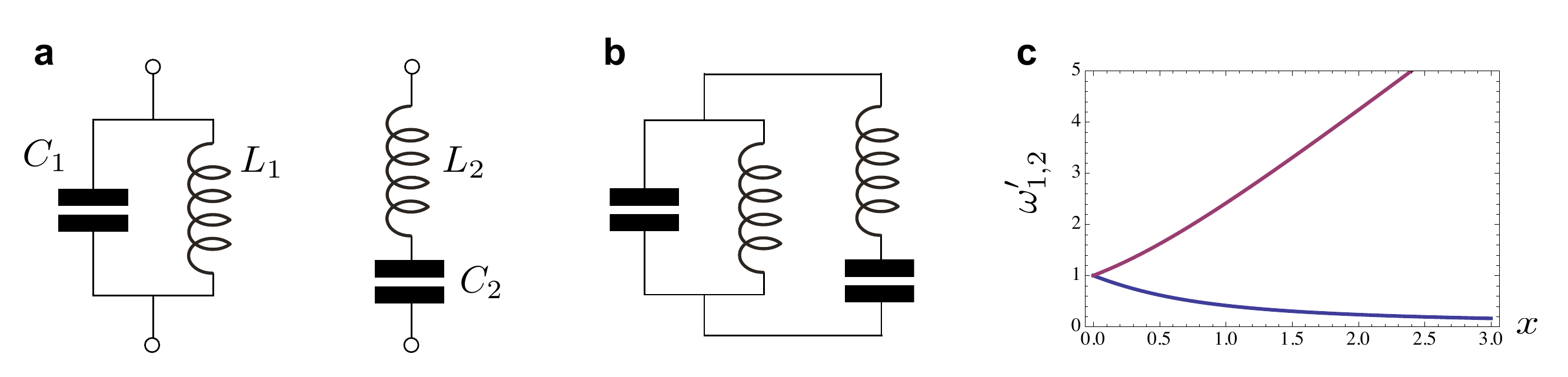}
	\caption{(a) A parallel (left) and a series (right) single-mode resonance $LC$ circuit. (b) Coupled circuits obtained by connecting parallel and series circuits with wires. (c) The two normal mode frequencies of the circuit shown in (b) for identical bare frequencies as a function of the dimensionless coupling parameter $x$ (see Table 1).}
	
	\label{fig:NormalModes}
\end{figure*}

A minimal circuit model illustrating circuit-circuit interaction consists of a parallel combination of a capacitor $C_1$ and an inductor $L_1$ shunted by a series combination of a capacitor $C_2$ and an inductor $L_2$ (Fig.~\ref{fig:NormalModes}a,b). Taking the two generalized fluxes $\phi_1$ and $\phi_2$  in the two capacitors~\cite{devoret1995quantum}, as the generalized coordinates, the circuit Hamiltonian is given by:
\begin{equation}\label{Hamiltonian-2-LC-circuits-Flux}
H_{\rm{flux}} = Q_1^2/2C_1 + Q_2^2/2C_2 + \phi_1^2/2L_1 + (\phi_1-\phi_2)^2/2L_2 .
\end{equation}
Here $Q_1$ and $Q_2$ are the conjugate momenta for $\phi_1$ and $\phi_2$, respectively. They obey the commutation relation $[\phi_i, Q_j] = i\hbar\delta_{ij}$. Physically, $Q_1$ and $Q_2$ are the displacement charges on the plates of the two capacitors $C_1$ and $C_2$.
 With this choice of circuit variables, the two $LC$-oscillators are inductively coupled via the term $\phi_1 \phi_2/L_2$. We thus call this choice of circuit variables a \textit{flux gauge}. Note that the coupling term does not contain any apparent small parameter. One can also write the Hamiltonian in a different gauge, by applying a unitary transformation $U = \mathrm{exp}{(-iQ_2\phi_1/\hbar)}$, which displaces $Q_1 \rightarrow Q_1+Q_2$ and $\phi_2 \rightarrow \phi_2 + \phi_1$, obtaining
\begin{equation}\label{Hamiltonian-2-LC-circuits-Charge}
H_{\rm{charge}}=(Q_1 + Q_2)^2/2C_1 + Q_2^2/2C_2 + \phi_1^2/2L_1 + \phi_2^2/2L_2.
\end{equation}  
In this so-called \textit{charge gauge} the circuit-circuit coupling is given by the capacitive term $Q_1 Q_2/C_1$ and again it does not apparently contain a small parameter. 

Already at this stage, it is interesting to note that the coupling between these two circuits cannot be classified as inductive or capacitive: clearly this is just a matter of a gauge choice which cannot affect the observables, such as frequencies of the normal modes. To understand the circuit-circuit coupling further we rewrite both Hamiltonians using creation and annihilation operators for the unperturbed modes,

%For the sake of clarity we will stick to the flux gauge in the remaining of this section, although all our %results can be safely reproduced in the charge gauge as well.

\begin{align}
& \phi_j = \sqrt{\frac{\hbar Z_j}{2}} (a_j + a_j^{\dagger}) \\
& Q_j = i\sqrt{\frac{\hbar}{2Z_j}} (a_j^{\dagger} - a_j),
\label{a}
\end{align}

where we have introduced mode impedances $Z_{1,2}$ according to the Table \ref{Table-coupled-LC}.
The resulting flux-gauge Hamiltonian is
\begin{equation}\label{Hamiltonian-LC}
H_{\rm{flux}}/\hbar = \omega_1 a_1^{\dagger}a_1 + \omega_2 a_2^{\dagger}a_2 - G (a_1^{\dagger}+a_1)(a_2^{\dagger}+a_2) +  D (a_1^{\dagger}+a_1)^2.
\end{equation}

The first term of the Hamiltonian~(\ref{Hamiltonian-LC}) is simply the harmonic oscillator energy of the parallel $LC$ circuit when its terminals are left open. Similarly, the second term is the energy of the series $LC$ circuit when its terminals are connected to each other. The $G$-term represents the linear coupling between the two $LC$-modes. The $D$-term renormalizes the bare parallel $LC$-mode frequency. Examination of the two constants (Table~\ref{Table-coupled-LC}) reveals an important dimensionless parameter $x$ that determines the strength of the coupling:
\begin{equation}
	\label{xxx}
	x = \frac12\sqrt{Z_1/Z_2}.
\end{equation}
The Hamiltonian (\ref{Hamiltonian-LC}) has actually precisely the same form that the Hopfield Hamiltonian  \cite{PhysRev.112.1555,PhysRevB.72.115303,PhysRevB.79.201303,Todorov2010}, and the $D$-term is the circuit analog of the diamagnetic "$A^2$-term" of the Hopfield model. Rewriting the charge-gauge Hamiltonian in terms of creation and annihilation operators  (see Table~\ref{Table-coupled-LC} for parameters) we obtain

\begin{equation}
H_{\textrm{charge}}/\hbar = \omega_1 a_1^{\dagger}a_1 +  \omega_2 a_2^{\dagger}a_2 +  \tilde{G} (a_1^{\dagger}-a_1)(a_2^{\dagger}-a_2) 
+  \tilde{D} (a_2^{\dagger}+a_2)^2.
\end{equation}

Let us assume for the sake of clarity that the two bare frequencies are identical, namely $\omega_1 = \omega_2 = \omega_0$. As it must be, in this case it is particularly apparent that the two Hamiltonians in both gauges have the same normal mode frequencies $\omega_{1,2}^{'}$. Exact diagonalization gives
\begin{equation}\label{Coupled-LC-frequencies}
\omega_{1,2}^{'}(x) = \omega_0 (\sqrt{1 + x^2} \pm x).
\end{equation}
In the weak coupling regime, $x\ll1$, we can ignore the $D$-term and recover a familiar normal mode splitting $\omega_{1,2}^{'}\approx\omega_0 \pm G$. In the opposite limit, $x\gg1$, the $D$-term must be taken into account, and we obtain that while the higher mode frequency continues to grow linearly with the coupling as $\omega_1^{'} \approx 2G$, the lower mode frequency now vanishes algebraically with $G$ as $\omega_2^{'} \approx \omega_0^2/2G$. Note that despite being small, $\omega_2{'}$ never reaches zero~(Fig. \ref{fig:NormalModes}c).

\begin{table}\label{Table-coupled-LC}
	\begin{tabular}{|c|c|c|c|c|c|c|c|c|}
		\hline  
		$\omega_1$ & $Z_1$ & $\omega_2$ & $Z_2$ & $x$ & $G$ & $D$ &$\tilde{G}$&$\tilde{D}$\\ 
		\hline 
		$1/\sqrt{L_1 C_1}$ & $\sqrt{L_1/C_1}$ &	$1/\sqrt{L_2 C_2}$  & $\sqrt{L_2/C_2}$ & $\frac12 \sqrt{Z_1/Z_2}$ & $\omega_1 \times x$ & $\omega_1 \times x^2$ &$\omega_2\times x$&$\omega_2 \times x^2$\\ 
		\hline 
	\end{tabular} 
	\caption{Parameters of the minimal model of coupling of two $LC$ circuits.}
\end{table}

We can now clearly see why it is so easy to go beyond weak coupling regime with circuits. Indeed, to make $x\sim 1$ it is sufficient to simply connect two circuits with identical inductances and capacitances! Adjusting circuit elements such that  $Z_1 \gg Z_2$ will enhance the coupling even further, making $x \gg 1$. In practice, the range of experimentally achievable circuit impedances is quite large. The lowest impedance of an electromagnetic oscillator is usually limited by the vacuum impedance $Z_{\textrm{vac}} = \sqrt{\mu_0/\epsilon_0}\approx 377~\Omega$, where $\epsilon_0$ and $\mu_0$ are the permittivity and permeability of the free space, respectively. On-chip microstructures often have impedance closer to $50~\Omega$ due to geometry and high dielectric constant substrates. To obtain high-impedance, one needs to increase the effective $\mu_0$. Impedance in excess of resistance quantum (for Cooper pairs) $R_Q = h/(2e)^2 \approx 6.5~k\Omega$ was achieved using a kinetic inductance of a Josephson junction chain while maintaining high coherence and showing no spurious resonances~\cite{Manucharyan113}. It is therefore relatively straightforward to achieve $x>5$.

We conclude this section with two important remarks. 

First, the two normal modes have a simple electrical engineer's interpretation in the ultra-strong coupling regime $x \gg 1$. The higher frequency mode can be understood as a resonance between $L_1$ and $C_2$ (current through $L_1$ and voltage across $C_2$ are neglected), while the lower frequency mode can be similarly understood as a resonance between $C_1$ and $L_2$ (current through $C_1$ and voltage across $L_2$ are neglected). In other words, the two circuits "exchange" their circuit elements such that the largest $L$ pairs up with the largest $C$ and the smallest $L$  pairs up with the smallest $C$, thus creating a large splitting in the two normal modes in the ultrastrong coupling regime.

Second, introducing a weak Kerr-like non-linearity to one of the oscillators will not modify the normal modes significantly, even if the coupling is strong. Indeed, let us assume that the inductance $L_1$ is slightly non-linear, i.e. there is a term quartic in $\phi_1$ in the energy. Such a non-linearity is typical for a transmon qubit~\cite{koch2007charge}. As remarked above, in the ultra-strong coupling regime, $x \gg 1$, the lower mode is predominantly confined to the inductance $L_1$ and capacitance $C_2$. Therefore, a weak non-linearity in $L_1$ would simply translate into a weak anharmonicity of the lower mode. Moreover, the higher mode, being predominantly confined to inductance $L_2$ and capacitance $C_1$ would be unaffected by such a  non-linearity in this (arguably very coarse) approximation. The only other type of non-linearity available in superconducting circuits is that associated with flux and charge tunneling  and is the main subject of this paper.

%\section{Circuit theory of a multilevel atom to a lumped-element $LC$ resonator for arbitrary coupling}
\section{Circuit theory of arbitrary strong atom-photon interactions}
\label{atom}
In this section we will describe the coupling of two types of circuit atoms - fluxonium and Cooper pair box to photons in a single mode $LC$-resonator~(Fig. \ref{fig:QubitCircuits}). 

\subsection{Fluxonium atom}

\label{fluxonium}
\begin{figure*}[t!]
	\centering
	\includegraphics[width=\linewidth]{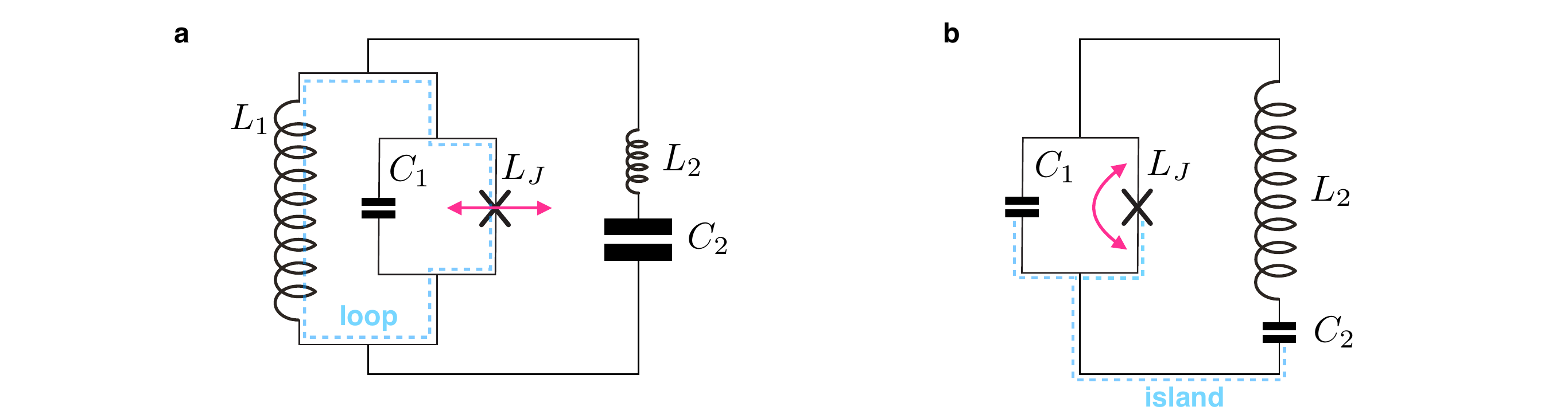}
	\caption{(a) The circuit scheme for the fluxonium-resonator system considered in Sec. \ref{fluxonium}, where the cross symbol represents a Josephson junction. The depicted loop is pierced by an external flux $\phi_{ext}$ (not shown). The main dynamics in this circuit is due to tunneling of flux in and out of the loop through the junction. (b) The circuit for the Cooper pair box case treated in Sec. \ref{CPB}. The depicted island can be additionally voltage-biased (not shown) to create an offset charge $Q_{\textrm{ext}}$. The main dynamics here is the tunneling of a charge in and out of the island across the junction.}
	
	\label{fig:QubitCircuits}
\end{figure*}

To maximize the fluxonium-photon coupling we keep the series $L_2 C_2$-oscillator from (Fig.~\ref{fig:NormalModes}b) and add a Josephson junction with the Josephson inductance $L_J$ in parallel with the inductance $L_1$ (Fig.\ref{fig:QubitCircuits}a). The loop formed by the junction and the linear inductance $L_1$ is pierced by the externally applied flux $\phi_{ext}$. Introducing the superconducting flux quantum $\Phi_0=h/2e$, we can write the uncoupled fluxonium atom Hamiltonian as

\begin{equation}\label{key}
H_{\textrm{fluxonium}} (\phi_1,Q_1) = Q_1^2/2C_1 + \phi_1^2/2L_1 - (\Phi_0/2\pi)^2/L_J \cos \left (2\pi \frac{\phi_1 - \phi_{ext}}{\Phi_0} \right ).
\end{equation}

Circuit parameters are to satisfy the following relations: $L_1 \gg L_J$ and $\sqrt{L_J/C_1}\sim R_Q$. The first condition ensures that there are multiple minima in the effective potential seen by the flux coordinate $\phi_1$; in practice this requires a rather large linear inductance which is achieved by a chain of order $100$ larger-area Josephson junctions \cite{Manucharyan113}. The second condition ensures that there is tunneling of the flux between the neighboring wells. The fluxonium-photon coupling Hamiltonian is more conveniently written in the flux gauge and is obtained by analogy with Eq.~(\ref{Hamiltonian-2-LC-circuits-Flux}):

\begin{equation}\label{eq:UltrastrongFluxonium}
H_{\textrm{fluxonium-photon}} = H_{\textrm{fluxonium}}(\phi_1,Q_1) + Q_2^2/2C_2 + (\phi_1-\phi_2\large)^2/2L_2.
\end{equation}

The superconducting loop is maximally frustrated at a specific value of external flux $\phi_{ext} = \Phi_0/2$. In that case, the atom's effective potential has a symmetric double-well shape consisting of two lowest degenerate local minima separated by approximately the flux quantum $\Phi_0$. The lowest transition frequency $\omega_{eg}$ from the bare ground state to the first excited state is due to the tunneling of the flux variable $\phi_1$ between these minima. Physically, this corresponds to a coherent tunneling of flux in and out of the superconducting loop and sometimes is referred to as a quantum phase-slip~\cite{manucharyan2012evidence}. The frequency of this tunneling transition can be much lower than the frequency of vibrations in the same well (often called plasma frequency), in which case the spectrum tends to be very anharmonic. It is then tempting to truncate $H_{fluxonium}$ to its lowest two states $|0\rangle$ and $|1\rangle$ according to the following substitution: 
\begin{align}
& \phi_1 \rightarrow \langle 0|\phi_1| 1\rangle \sigma_x\\
& H_{\textrm{fluxonium}}(\phi_1,Q_1) \rightarrow \frac12 \hbar\omega_{01} \sigma_z.
\end{align}
By making this substitution into the atom-photon Hamiltonian (\ref{eq:UltrastrongFluxonium}), relabeling $\omega_2 \rightarrow \omega_r$, $a_1 \rightarrow a$, and $Z_2 \rightarrow Z_r$ and using the fact that $\sigma_x^2 = 1$ we indeed obtain an effective quantum Rabi model with the dimensionless coupling constant (see Eq. (\ref{Rabi}))
\begin{equation}
\label{Eq: gFlux}
\frac{g}{\omega_r} =\frac{2 \langle 0|\phi_1|1\rangle}{\Phi_0}\times x
\end{equation}
where $x=\frac12\sqrt{\pi R_Q/Z_r}$ is independent of the atom spectral details. For a typical fluxonium circuit $\langle 0|\phi_1|1\rangle \approx \Phi_0/2$ implying $g/\omega_r \approx x$. Therefore, a coupling strength $ x > 1$ can be readily implemented using $Z_r \sim 50~\Omega$  resonance circuits~\cite{yoshihara2016superconducting}. Comparing Eq.~(\ref{Eq: gFlux}) to the expression for $x$ for linear circuits (see Table~\ref{Table-coupled-LC}) we can formally assign to fluxonium circuit the effective impedance $Z_{\textrm{fluxonium}} \approx \pi R_Q$. 

Let us also note that moving away from the maximal frustration point (sometimes referred to as flux degeneracy point) would introduce an additional transverse term to the truncated atom's Hamiltonian proportional to $(\phi_{ext}-\Phi_0/2)\sigma_x$. This term would steer the system away from the conventional quantum Rabi model and prevent formation of entangled atom-photon ground state. Therefore maximal frustration of the flux variable is crucial  for implementing the quantum Rabi model.

\subsection{Cooper pair box atom}
\label{CPB}
Maximally strong coupling of a Cooper pair box circuit to a resonator is also achieved by shunting it with a series $LC$-circuit. Formally this is equivalent to removing the linear inductance $L_1$ from the fluxonium circuit~(Fig.~\ref{fig:QubitCircuits}b). The resulting circuit contains an "island", i.e. a superconducting region, whose total charge is quantized and can only change by a tunneling of a Cooper pair. This island can be additionally voltage-biased (not shown) to create an offset charge $Q_{\textrm{ext}}$. The Hamiltonian of this system, called a Cooper pair box (CPB) atom, is given by

\begin{equation}
\label{Eq: CPB}
H_{\textrm{CPB}}(\phi_1, Q_1) = (Q_1 + Q_{ext})^2/2C_1 - E_J \cos(2e\phi_1/\hbar).
\end{equation}
The operator $Q_1$ has integer eigenvalues in units of $2e$ and its conjugate flux $\phi_1$ is a compact variable defined on the interval $[0,h/2e)$. Consequently, the flux-charge commutation relation here changes to $\exp(iq\phi_1/h) Q_1 \exp(-iq\phi_1/h) = Q_1 +q$. The atomic spectrum can be understood as that of a particle in a periodic cosinus potential with a quasi-momentum given by the offset charge $Q_{ext}$~\cite{bouchiat1998quantum}. The atom-photon coupling in this case is more conveniently written in the charge gauge, by analogy with Eq.~(\ref{Hamiltonian-2-LC-circuits-Charge}):
\begin{equation}
\label{Eq: UltrastrongCPB}
H_{\textrm{CPB-photon}} = H_{\textrm{CPB}}(\phi_1,Q_1+Q_2) + Q_2^2/2C_2 + \phi_2^2/2L_2.
\end{equation}

In this gauge the $Q_2^2$-term coming from $H_{CBP}(\phi_1, Q_1 + Q_2)$ strongly renormalizes the resonator frequency. It turns out that the analysis appears more intuitive if we take into account this renormalization explicitly. Physically, this corresponds to replacing the resonator capacitance $C_2$ with a parallel combination $C_1||C_2 = C_1 C_2/(C_1 + C_2)$. This is particularly evident in the limit $E_J \rightarrow 0$. The photon annihilation operator $a$ is defined via the renormalized resonator frequency $\omega_r = 1/\sqrt{L_2 C_1 || C_2}$ and the renormalized resonator impedance $Z_r = \sqrt{L_2/C_1||C_2}$, resulting in the following CPB-photon coupling Hamiltonian

\begin{equation}\label{Eq: CPB-LC}
H_{\textrm{CPB-photon}} = H_{CPB}(\phi_1, Q_1) + \hbar\omega_r a^{\dagger}a + \hbar \omega_r  \frac{C_2}{C_1 + C_2} \sqrt{\pi Z_r/R_Q} \times i(a^{\dagger}-a)\frac{Q_1}{2e}.
\end{equation}

In a Cooper pair box, the charge variable is maximally frustrated at $Q_{ext} = -e$. Similarly to the case of fluxonium, it is crucial to operate the Cooper pair box at maximal charge frustration (charge degeneracy point) for obtaining the quantum Rabi model. In this case the charging energy of the Cooper pair box is degenerate for $Q/2e = 0,1$, i.e. for the absence or presence  of an extra Cooper pair on the island. The Josephson term lifts the degeneracy by tunneling of Cooper pairs. Truncating the Hamiltonian~(\ref{Eq: CPB-LC}) to the two lowest atomic states $|0 \rangle$ and $|1 \rangle$, we get
\begin{eqnarray}
Q_1 \rightarrow \langle 0|Q_1|1\rangle\sigma_z,\\
\cos(2e \phi_1/\hbar) \rightarrow  \frac{\sigma_x}{2}.
\end{eqnarray}
This truncation gives a quantum Rabi model (see Eq. \ref{Rabi}) with the dimensionless coupling constant $g$ 
\begin{equation}
\frac{g}{\omega_r} = \frac{C_2}{C_1 + C_2} \times \frac{\langle 0|Q_1|1\rangle}{e}\times x,
\label{CPB_gRabi}
\end{equation}
where this time 
\begin{equation}
x = \frac12 \sqrt{\pi Z_r/R_q}.
\label{CPB_x}
\end{equation}
For typical CPB parameters, $ \langle 0|Q_1|1\rangle/e \sim 1$, so as long as we choose $C_1 \ll C_2$, we get a particularly simple result $g/\omega_r \approx x$. To enhance coupling, we need a series circuit with a large inductance $L_2$, such that $Z_r \approx \sqrt{L_2/C_1} \sim R_Q$. Interestingly, capacitance $C_2$ is irrelevant as long as $C_1 \ll C_2$. This is a consequence of properly taking into account the "$A^2$"-like term in deriving Eq.~(\ref{Eq: UltrastrongCPB}). Taking the typical small junction capacitance value $C_1 \approx 1~\rm{fF}$ and the largest demonstrated linear inductance $L_2 \approx 300~\rm{nH}$, we get $Z_r \approx 17~k\Omega$~\cite{Manucharyan113} and $x \approx 1.5$, sufficient to reach the ultrastrong coupling regime with a charge qubit.

\begin{table}[t!]
\begin{tabular}{|c|c|c|c|c|}
	\hline 
	Atom/Parameter & $\omega_r$ & $Z_r$ & $g/\omega_r$ & $x$ \\ 
	\hline 
	Fluxonium & $1/\sqrt{L_2 C_2}$ & $\sqrt{L_2/C_2}$ & $\cfrac{2 \langle e| \phi_1 | g \rangle}{\Phi_0}\times x$ & $\frac12 \sqrt{\pi R_Q/Z_r}$ \\ 
	\hline 
	Cooper pair box & $1/\sqrt{L_2 C_1 C_2/(C_1+C_2)}$ & $\sqrt{L_2 (C_1 + C_2)/C_1 C_2}$ & $\cfrac{C_2}{C_1+C_2}\cfrac{\langle e|Q_1|g\rangle}{e}\times x $& $\frac12 \sqrt{\pi Z_r/R_Q}$ \\ 
	\hline 
\end{tabular}
\caption{Parameters of the quantum Rabi model as a result of the two-level truncation for the two types of circuit atoms considered in the text. They depend on the circuit inductances/capacitances (Fig.~\ref{fig:QubitCircuits}) and on the flux/charge matrix elements between the bare atom ground state and the first excited state.} 
	\label{Table: QRM}
\end{table}

\section{Numerical diagonalization of multilevel circuit Hamiltonians}
\label{discussion}

The two-level truncation for both circuit atoms yielded a quantum Rabi model with $g/\omega_r >1$ for realistic circuit parameters. Are we allowed to perform such a truncation? What is the role of higher atomic levels in the ultrastrong coupling regime? To answer this question we compare the results of the numerical simulation of the Hamiltonians (\ref{eq:UltrastrongFluxonium},\ref{Eq: UltrastrongCPB}) keeping over $100$ atom and resonator levels to the prediction of their truncated quantum Rabi model implementations. It is convenient in our simulations to describe circuit parameters using energy scales given by $E_C = e^2/2C$ for a capacitance $C$, $E_L = (\Phi_0/2\pi)^2/L$ for an inductance $L$, and $E_J = (\Phi_0/2\pi)^2/L_J$ for the Josephson junction.  In presenting results, we will measure these energies and circuit transition frequencies in units of $\rm{GHz}$.

\subsection{Frequency spectra for the fluxonium-resonator circuit}

The three numbers $E_J$, $E_{L_1}$, and $E_{C_1}$, together with the external flux $\phi_{ext}$ define the fluxonium transition spectrum. We chose a set $(E_J,E_{C_1},E_{L_1})=(5,5,0.5)~\rm{GHz}$ which is experimentally realistic and yields a highly anharmonic spectrum with $\omega_{01}/2\pi \approx 2.47$ and $\omega_{02} > 3\times \omega_{01}$. We consider a conventional resonant case ($\omega_r/2\pi = 2.47$) and two off-resonant cases ($\omega_r/2\pi = 8, 20$). The case $\omega_r/2\pi = 8$ is special because then $\omega_{12} \approx \omega_r$, while $\omega_r/2\pi = 20$ represent a high-frequency limit of $\omega_r \gg \omega_{01}$.

\begin{figure*}[h]
	\centering
	\includegraphics[width=\linewidth]{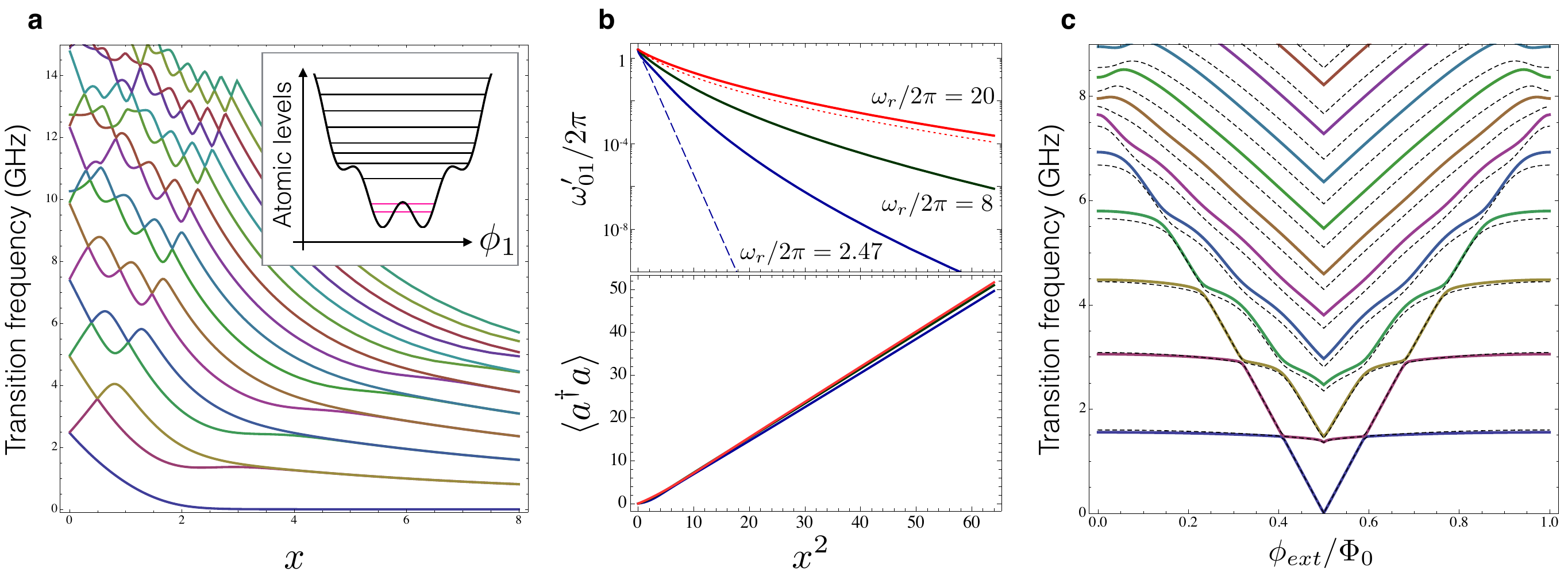}
	\caption{(a) Transition spectrum of the Hamiltonian~(\ref{eq:UltrastrongFluxonium}) including over $100$ atom and 100 photon levels for $\omega_r = \omega_{01}=2.47$. The fluxonium atom parameters are $(E_J, E_{C_1}, E_{L_1}) = (5,5,0.5)$. Inset shows the spectrum of the bare (uncoupled) atom levels within the flux-dependent potential. (b-top) Frequency splitting $\omega_{01}'$ between the ground state and the first excited state for three bare photon frequencies, $\omega_r = 2.47$ (blue), $\omega_r = 8$ (dark green) and $\omega_r=20$ (red). Dashed blue line is the result of the quantum Rabi model, dotted red line is the result with $L_2 =0$ (a capacitively shunted fluxonium model). (b-bottom) Expectation value of the photon number in the ground state for the same three resonator frequencies. (c) Transition spectrum of Hamiltonian~(\ref{eq:UltrastrongFluxonium}) as a function of the external flux for $x=4$ (solid lines) and spectrum of the uncoupled fluxonium with renormalized parameters (dashed lines).}
	\label{fig:FluxoniumUSC}
\end{figure*}

Our numerical simulations show that while the spectrum of a fluxonium coupled to a single-mode photon is indeed qualitatively similar to that of the quantum Rabi model, there are several striking differences (Fig.~\ref{fig:FluxoniumUSC}). The features similar to the quantum Rabi model are: (i) the spectrum at $x \gg 1$ reduces to a harmonic ladder of nearly degenerate pairs of levels (Fig.~\ref{fig:FluxoniumUSC}a) and (ii) the expectation of the number of photons in the resonator in the ground state grows as $x^2$, being insensitive to the value of the resonator frequency (Fig.~\ref{fig:FluxoniumUSC}b - bottom). The key discrepancy from the quantum Rabi model are: (I) the frequency gap between nearly degenerate level doublets reduces algebraically with $x$ as opposed to the prediction of being $x$-independent for $x \gg 1$ and (II) the even-odd ground state splitting scales sub-exponentially with $x^2$ (or, equivalently, sub-exponentially with the number of ground state photons). Furthermore, the ground-states splitting goes up by orders of magnitude when the bare resonator frequency is a few times larger than the bare atom's transition frequency (Fig.~\ref{fig:FluxoniumUSC}b). 

We now offer a physical interpretation to our numerical results. We first discuss the quadratic growth of the ground-state photon number as a function of the dimensionless coupling $x$. It can be estimated from the amount of displacement of the oscillator by the atom. If we disregard tunneling, the oscillator is displaced approximately by $\pm \Phi_0/2$. This corresponds to the coherent state amplitude $\alpha = \Phi_0/(\langle0|\phi_2^2|0\rangle)^{1/2}$, which happens to equal exactly $x$, such that the ground state photon number is given by $\langle a^{\dagger}a\rangle = x^2$. Quantum tunneling of the atomic coordinate merely entangles the two oppositely displaced oscillator states and does affect the ground-state photon number.

Next, we turn to the spectral features of the Hamiltonian~(\ref{eq:UltrastrongFluxonium}). Why does the scaling of the lowest transition frequency deviate so much from the predictions of the quantum Rabi model? To understand this, let us consider an atomic transition between states $0$ and $1$ and calculate to second order in $x$ its shift $\chi_{01}$ due to the interaction with a cavity. We will also assume no resonance conditions, i.e.   $\omega_{0i},\omega_{1j} \neq \omega_r$.  The answer is given by~\cite{zhu2013circuit} 

\begin{equation}
\chi_{01} = x^2 \omega_r^2 \sum_{i\neq0} \frac{2\omega_{0i}}{\omega^2_{0i}-\omega^2_r}|\langle 0|2\phi_1/\Phi_0|i\rangle|^2 - x^2 \omega_r^2 \sum_{j\neq1} \frac{2\omega_{1j}}{\omega^2_{1j}-\omega^2_r}|\langle 1|2\phi_1/\Phi_0|j\rangle|^2 .
\end{equation}
This perturbative expression for $\chi_{01}$ can differ greatly from its 2-level truncated version (the terms $i=1, j=0$) if there are transitions from either state $0$ or $1$ to higher atomic states with non-zero matrix elements and with frequencies close to $\omega_r$. In other words, already in second order in $x$, the shift $\chi_{01}$ can be large even if $\langle 0|2\phi_1/\Phi_0|1\rangle| =0$ and $\omega_{01}\ll \omega_r$. Clearly truncation to the two states is even more dangerous for $x\gg1$, no matter how anharmonic the atomic spectrum is. 

Although perturbation theory fails at $x\gg1$ we can still make sense of the low-energy spectrum in the high photon frequency limit $\omega_r \gg \omega_{01}$. In this limit it is tempting to just ignore a voltage drop across the inductance $L_2$ thus reducing the effect of the resonator to simply shunting the qubit with the capacitance $C_2$; mind that to get $x\gg1$ one needs $ C_2 \gg C_1$. As a result, flux tunneling is suppressed since the capacitance plays the role of a mass for the tunneling of the flux coordinate $\phi_1$. This suppression is similar to the familiar suppression of macroscopic quantum tunneling of phase in large-capacitance Josephson junctions~\cite{clarke1988quantum}. The WKB approximation predicts that the tunnel splitting will scale exponential in $\sqrt{C_2}$ which is proportional to $x$ if we keep the resonator frequency constant while increasing the coupling. Our numerical calculation, where we ignore the inductance $L_2$ (Fig.~\ref{fig:FluxoniumUSC}b, red dotted line) agrees well with the full numerical simulation using $\omega_r = 20$ and $\omega_{01} = 2.47$ (Fig.~\ref{fig:FluxoniumUSC}b, red solid line).

Finally, we observe that the energy spectrum of a quantum Rabi model at large couplings is qualitatively very similar to the spectrum of a single fluxonium atom with an enhanced ratio $E_J/E_{C_1}$ (large barrier, heavy mass). The flux particle can semi-classically vibrate in either left or right wells which creates the 2-fold degeneracy in the spectrum. Exponentially small tunneling through the barrier weakly lifts the degeneracy. In other words, the main effect of ultrastrong coupling of a fluxonium (or a flux qubit) to a cavity, from a low-energy spectroscopic point of view, reduces to a simple suppression of coherent flux tunneling. To illustrate this point, we fix $x = 4$, deep in the non-perturbative coupling regime, and simulate the spectrum of Eq.~(\ref{eq:UltrastrongFluxonium}) as a function of flux $\phi_{ext}$. The resulting spectrum (Fig.~\ref{fig:FluxoniumUSC}c, solid lines), when restricted to the lowest few states is remarkably similar to that of a bare fluxonium with renormalized parameters (Fig.~\ref{fig:FluxoniumUSC}c, dashed lines). The difference between the two models can only be found in quantitative discrepancy of the higher-frequency transitions.

\subsection{Frequency spectra for the Cooper pair box circuit}

Prior to presenting our numerical results on a CPB, let us first remark on our choice of varying $x$. Because the bare photon frequency $\omega_r$ in case of the CPB-photon coupling depends on the junction's capacitance~(see Table~\ref{Table: QRM}), it is impractical to study the variation of the coupling constant $x$ while keeping $\omega_r$ constant. Instead we chose to increase $x$ by increasing $L_2$, which would simultaneously reduce the bare photon frequency as $1/x^2$. While this choice makes the spectrum look strange at $x\sim 0$ due to divergence of the photon frequency, it is much easier to interpret it at $x>1$ which is the main goal of our study. Moreover, it is most practical in an experiment to vary resonator's inductance while keeping the rest of the circuit constant. 

\begin{figure*}[h]
	\centering
	\includegraphics[width=\linewidth]{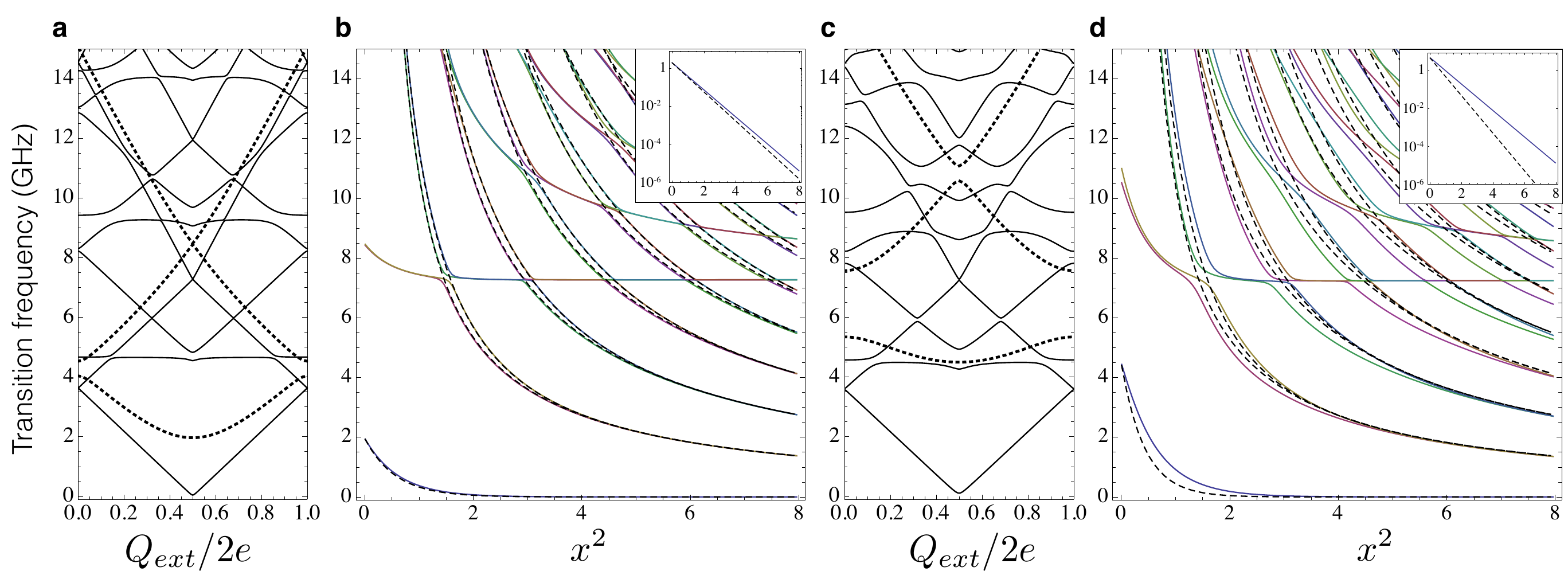}
	\caption{\textbf{(a)} Transition spectrum of the bare Cooper pair box (Eq.~\ref{Eq: CPB}) [dotted lines] and the coupled Hamiltonian (Eq.~\ref{Eq: UltrastrongCPB}) [solid lines] as a function of the offset charge $Q_{ext}$ for circuit parameters $(E_J, E_{C_1}, E_{L_2}, E_{C_2}) = (2,10,0.25,1)$. \textbf{(b)} Transition spectrum of the coupled Hamiltonian [solid lines] and that of its truncated 2-level approximation (Eq.~\ref{Rabi} and Eq.~\ref{CPB_gRabi}) [dashed lines] for $Q_{ext}=e$ as a function of dimensionless coupling $x^2$, defined by Eq.~\ref{CPB_x}.  Note that parameter $x$ was varied by increasing the inductance $L_2$ while keeping other circuit parameters fixed; this results in $x$-dependent bare photon frequency, which diverges at $x=0$.  Inset shows the lowest transition frequency on the log scale for both models. Inset shows a zoom-in on the lowest transition with a log scale for the frequency. \textbf{(c)} Same as (a) for $E_J = 5$. \textbf{(d)} Same as (b) for $E_J = 5$.}
	\label{fig: CPBUSC}
\end{figure*}

Our results for the Cooper pair box are summarized in Fig.~(\ref{fig: CPBUSC}). In contrast to the fluxonium case, here the full model shows very minor discrepancy from the truncated quantum Rabi model. We have fixed the capacitances $C_1$ and $C_2$ such that $E_{C_1} = 10$, typical for a small-capacitance tunnel junction, and $E_{C_2}=1$ in order to respect the condition $C_1 \ll C_2$. We varied the coupling $x$ by varying $L_2$ (see Table~\ref{Table: QRM} and discussion in the text): this also varies the bare photon frequency, but we think that this choice is the most experimentally relevant. We show the results for two different values of the Josephson energy: $E_J =2$ (Fig.~\ref{fig: CPBUSC}a,b), representing stronger anharmonicity and $E_J=5$ (Fig.~\ref{fig: CPBUSC}c,d), representing weaker anharmonicity.

Our main finding is that the low-energy spectrum of the coupled atom-photon system can essentially be understood as a suppression of $E_J$ exponentially in $x^2$ irrespectively on the amount of anharmonicity of the bare atom. This is particularly evident in the case $(E_J, E_{C_1}) =(2,10)$  (Fig.~\ref{fig: CPBUSC}a), where one can recognize spectral lines of a Cooper pair box with $E_J \sim 0$. Higher atomic transitions almost do not interact with the rest of the spectrum (Fig.~\ref{fig: CPBUSC}b). Scaling of the lowest transition frequency with $x^2$ essentially coincides with the quantum Rabi model's result over many orders of magnitude. The effect of higher CPB levels is more prominent in case of the less anharmonic atom $(E_J, E_{C_1}) = (5,10)$, where the lowest bare atomic transition is almost charge-insensitive, like in the transmon qubit. Naively, one would expect a drastic departure from the quantum Rabi model, because the second atomic transition frequency is very close to the first one (Fig.~\ref{fig: CPBUSC}c - dashed lines). Nevertheless, the lowest transition frequency still scales exponentially with $x^2$ and the deviation from the quantum Rabi model is only in the renormalization of the coupling constant by a factor of order unity (Fig.~\ref{fig: CPBUSC}b - inset).

Such \textit{insensitivity} of the Cooper pair box circuit to its higher levels, even when $x \gg 1$ and when $E_J/E_{C_1}$ is not small (approaching the trasmon regime) is likely due to a rapid suppression of the charge transition dipole matrix element with the level number. The main spectral feature of the ultrastrong coupling for a charge qubit - the exponential in $x^2$ suppression of $E_J$  at $x \gg 1$ - can be qualitatively understood as dressing of a Cooper pair with a cloud of virtual photons, whose size grows with the coupling constant. As a result of this dressing, coherent Cooper pair tunneling is suppressed when $x\gg 1$. Indeed, the width of the ground state wavefunction of the uncoupled oscillator is given by $\langle 0|(Q_2/2e)^2|1\rangle^{1/2} \sim 1/x$ in the charge representation. Since tunneling of a Cooper pair shifts the oscillator's charge variable $Q_2/2e$ by a unity, this process will be suppressed exponentially in $x^2$ due to a small Frank-Condon overlap of the oscillator states before and after the tunneling event. Thus, unlike the suppression of flux tunneling in fluxonium, renormalization of $E_J$ here cannot be explained by a simple capacitive (or inductive) loading of the junction. In fact, the physics of ultrastrong coupling of a charge qubit to a single-mode photon appears to be very similar to dynamical Coulomb blockade, which describes the non-linearity of the inelastic charge-transport in tunnel junctions coupled to a high-impedance leads~\cite{tunneling1992h}.

\subsection{Entanglement spectrum of the ground state}
\label{entanglement}

In the quantum Rabi model, when $g \gg \omega_r =  \omega_{eg}$ the ground state tends asymptotically
to the Schroedinger cat state:
\begin{equation}
\vert G_{\rm Q-Rabi} \rangle \simeq {\mathcal N} \left (  \vert \alpha \rangle \vert \leftarrow \rangle - \vert -\alpha \rangle \vert \rightarrow  \right) \rangle
\label{cat}
\end{equation}
where $\vert \alpha \rangle$ is a coherent state for the resonator with amplitude $\alpha = g/\omega_r$,  $\sigma_x \vert \rightarrow \rangle  = \vert \rightarrow \rangle$ and $\sigma_x \vert \leftarrow \rangle  = - \vert \leftarrow \rangle$. The coefficient ${\mathcal N}$ is the normalization constant.

To characterize the entanglement between atom and resonator in the ground state, a very useful quantity is the entanglement spectrum \cite{PhysRevLett.101.010504}.
Let us call $\vert G \rangle$ the ground state of a bipartite system where $A$ and $B$ are the two partitions.
We can define a pure state density-matrix $\rho = \vert G \rangle \langle G \vert$ for the ground state.
By partial trace with respect to subsystem $B$, we get the reduced density matrix for the subsystem $A$, namely $\rho_A = {\rm Tr}_B (\rho)$.
%Analogously, $\rho_B = {\rm Tr}_A (\rho)$. The entanglement entropy is the Von Neumann entropy of the reduced density matrix, namely:
%\begin{equation}
%S_{\rm ent} = {\rm Tr}_A (-\rho_A {\rm ln} \rho_A) = {\rm Tr}_B (-\rho_B {\rm ln} \rho_B) .
%\end{equation}
The reduced density matrix is a hermitian operator, hence it can be diagonalized by an orthonormal basis of eigenstates $\vert \Psi_r^{(A)}\rangle $ with eigenvalues 
$0 \leq p_r \leq 1$ such that $\sum_r p_r = 1$, namely
\begin{equation}
\rho_A = \sum_r p_r \vert \Psi_r^{(A)}\rangle \langle \Psi_r^{(A)}\vert . 
\end{equation}
The entanglement spectrum is just  the set of probabilities $\{ p_r \}_r$, i.e., the spectrum of eigenvalues of the reduced density matrix. 
For a separable state (no entanglement), the reduced density matrix corresponds to a pure state, hence the entanglement spectrum contains the values $p_1 = 1$ and $p_{r \neq 1} = 0$. 
In the quantum Rabi model, for $\vert \alpha \vert \gg 1$, the entanglement spectrum associated to the ground state in Eq. (\ref{cat}) is $\{ p_1 = 1/2, p_2 = 1/2 \}$.

\begin{figure*}[t!]
	\centering
	\includegraphics[width=1.1 \linewidth]{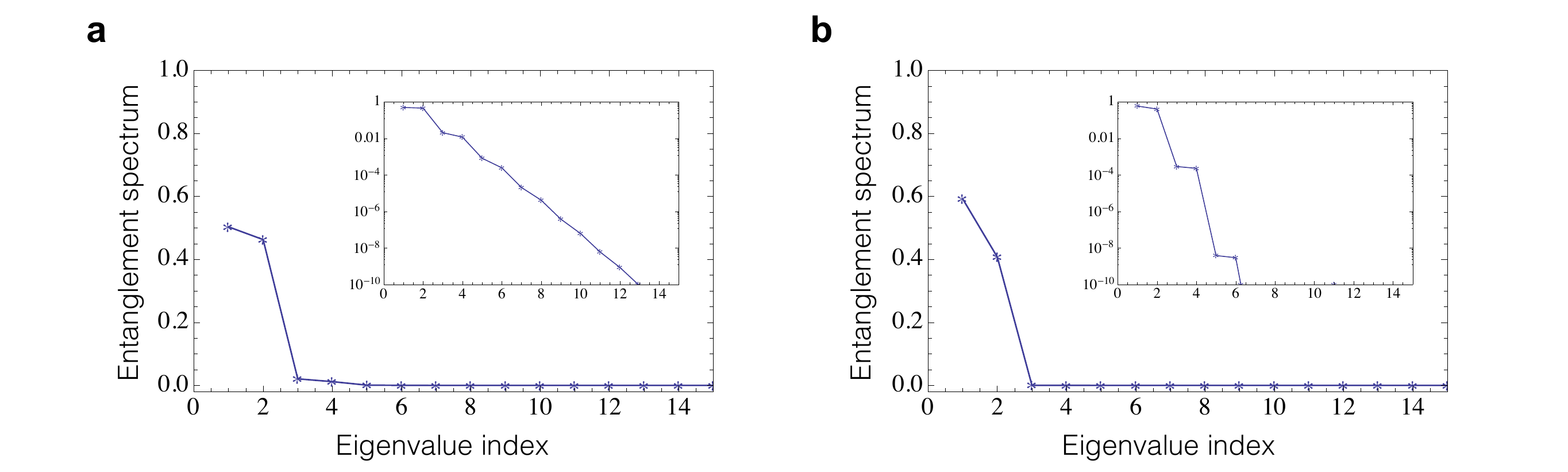}
	\caption{ Entanglement spectrum for the atom-resonator system. (a) Results for a fluxonium circuit (parameters as in Fig.~\ref{fig:FluxoniumUSC}, $x=4$). (b) Results for a Cooper pair box (same parameters as in Fig.~\ref{fig: CPBUSC}c).}
	\label{fig:entanglement}
\end{figure*}

In Fig. \ref{fig:entanglement} we show results for the entanglement spectrum for the ground state of  the fluxonium circuit (left panel) and for the Cooper pair box case (right panel). 
For both circuits we have considered a large dimensionless coupling $x = 4$. In both cases, the entanglement spectrum is dominated by two nearly equal eigenvalues, similarly to the quantum Rabi model case. In the inset, the eigenvalues are plotted in log scale to see the other and much smaller eigenvalues $p_{r \geq 3} \ll 1$. It is clear than in the fluxonium case there are some minor deviations with respect to the simple entanglement spectrum of the quantum Rabi model. Such deviations are even smaller for the Cooper pair box as the other eigenvalues are smaller by several orders of magnitude.  Indeed, the Cooper pair box provides a very close match to the ideal quantum Rabi model also for very large values of the dimensionless coupling $x$. 

Which is the physical interpretation of such simple entanglement spectrum with two dominant eigenvalues for the multilevel circuit atom? If we partially trace the pure  ground state density-matrix with respect to the resonator degrees of freedom, one obtains a reduced density matrix $\rho_{at}$ for the atomic system. A mixed density matrix $\rho_{at}$ implies entanglement between the atom and the resonator. In the quantum Rabi model, the dimension of the atomic Hilbert space is two, so trivially there are only two possible eigenvalues and two corresponding atomic eigenstates. Maximum entanglement is achieved when $p_1 = p_2 = 1/2$. In a multilevel atom instead it is not trivial at all that the entanglement spectrum is dominated by {\it only two} eigenvalues. The corresponding eigenvectors do define an effective two-level description, although they are not at all the bare atomic states. We have verified (not shown) that the most probable eigenstates of the atomic reduced density matrix  $\rho_{at}$ (corresponding to the two dominant eigenvalues $p_1 \approx p_2 \approx 1/2 $ in Fig. \ref{fig:entanglement}) are linear superpositions of {\it many} bare atomic states.

\section{Summary and outlook}
\label{conclusions}

In conclusion, we have investigated theoretically two classes of superconducting circuits where an artificial atom -- a fluxonium and a Cooper pair box -- is coupled to a lumped-element $LC$-resonator. Our studies of the fluxonium atom cover qualitatively the case of the junction flux qubit, while the studies of a Cooper pair box included a relatively large $E_J/E_C$ ratio characteristic of a transmon regime. We first derived and quantized circuit Hamiltonians in various gauges following Ref.~\cite{devoret1995quantum} and fully taking into account "$A^2$"-like terms. To achieve the regime $g/\omega_r >1$ we found that a low-impedance resonator is required in case of flux tunneling, which is consistent with the previous studies~\cite{ANDP:ANDP200710261, PhysRevA.80.032109}, and a high-impedance resonator is required for charge tunneling. The impedance scale is defined by the resistance quantum for Cooper pairs $R_Q = h/(2e)^2$. In both cases a value for the dimensionless coupling constant $x\sim\text{10}$ can be achieved with the existing technology. We then numerically explored spectral features of our Hamiltonians taking into account a large number of bare atom and photon states, a study that was not previously performed for either circuit. 

Despite significant deviation of both circuits from two-level systems, many essential aspects of the quantum Rabi model were recovered in the limit $g/\omega_r \gg 1$. Those include the entangled atom-photon structure of the ground state with a lifting of the ground state degeneracy that rapidly vanishes as $g/\omega_r$ grows and becomes larger than unity, and a nearly harmonic excitation spectrum. The ground state degeneracy is lifted exponentially  in $(g/\omega_r)^2$ in case of the Cooper pair box circuit, which coincides with the quantum Rabi model's prediction, while significant order of magnitude corrections were found for the fluxonium. The deviation is particularly notable when the bare photon frequency $\omega_r$ matches some of the higher transitions of the bare atom, such as $\omega_{02}$ or $\omega_{12}$.

A physical interpretation of the regime $g/\omega_r \gg 1$ for both flux and charge tunneling circuits is worth a separate note. In the case of fluxonium (or a flux qubit), the main effect is the suppression of flux tunneling. The origin of suppression is evident in the high photon frequency case, $\omega_r \gg \omega_{01}$: one can simply ignore the resonator inductance and treat the overcoupled atom-oscillator system as a fluxonium (or flux qubit) shunted by the large capacitance of the low-impedance $LC$-circuit. As a result, the flux tunneling is suppressed by analogy with the suppression of macroscopic quantum tunneling of phase in large-capacitance Josephson junctions~\cite{clarke1988quantum}. In the case of the Cooper pair box circuit, the main effect is the suppression of Cooper pair tunneling, irrespective of the amount of anharmonicity of the bare circuit. Here, the suppression is of the Frank-Condon type and cannot be explained by a  renormalization of the linear circuit elements. The suppression of $E_J$ by a high-impedance resonator environment draws analogies to the dynamical Coulomb blockade of charge transport in small-capacitance tunnel junction~\cite{tunneling1992h}. 

Finally, let us emphasize a striking property of both types of circuit atoms: as a long charge/flux variables are maximally frustrated (circuits are biased at their respective degeneracy points), all low-energy spectral features qualitatively match those of the quantum Rabi model, as long as $g/\omega_r \gg 1$. At a first sight this behavior appear surprising because the bare spectrum of circuit atoms can be very far from that of a two-level system. For instance, a CPB could be taken in the transmon regime of $E_J/E_C \gg 1$ where it behaves as weakly non-linear oscillator at low energies. Likewise, fluxonium could be taken with a very shallow double well potential. We believe the explanation lies in the underlying semi-classical "two-levelness" of both circuits when they are maximally frustrated at their degeneracy points. For instance, dressing of flux by photons makes this variable very "heavy" and it is left to sample only the bottom of the two wells, no matter how shallow the wells are, thereby restoring the two-level behavior. This effective dynamics occurs at a significantly reduced energy scale and the two lowest states consist of a large number of bare atom and photon states. In other words, it is not the strong anharmonicity of the flux qubit that is required to observe quantum Rabi dynamics at $g/\omega_r \gg 1$. Instead the key is the symmetric double-well potential seen by the flux degree of freedom, which is essentially a classical property of this circuit. Similar reasoning holds for the charge states of a superconducting island. 

Our theoretical results seem encouraging for further experimental exploration of the physics of the quantum Rabi model for large couplings. Although spectral features of ultrastrong qubit-oscillator system are relatively straightforward to observe~\cite{yoshihara2016superconducting}, a study of coherent quantum Rabi dynamics might be more challenging. Most notably, at large $g/\omega_r$, both charge and flux circuits become first-order sensitive to charge and flux noise respectively, which would limit coherence times to only few nanoseconds for charge and flux qubits. This is partially mitigated in case of fluxonium, where a large loop inductance results in first-order flux-noise decoherence in excess of $1~\mu s$~\cite{manucharyan2012evidence, kou2016fluxonium}. Non-destructive measurement of the non-trivial ground state properties can be done for example using an ancillary qubit \cite{PhysRevLett.114.183601,FelicettiSciRep}. An efficient release of ground state photons (the so-called radiation from the vacuum) can occur only if the system parameters are changed in a non-adiabatic fashion \cite{PhysRevA.74.033811,PhysRevLett.98.103602,PhysRevA.80.053810,Gunter2009,Peropadre10}, which can probably be achieved using a fast flux/charge modulation in the circuits considered here.

\acknowledgements{V.E.M. acknowledges support from the US National Science Foundation (DMR-1455261) and from the Alfred P. Sloan Research Fellowship (FG-2015-66004).  A.B. acknowledges support from the Air Force Office of Scientific Research. C. C. acknowledges support from ERC (via the Consolidator Grant "CORPHO" No. 616233).}

% placement option to break a long table (with less control than 
% \begin{table}%[H] add [H] placement to break table across pages
% \begin{ruledtabular}
% Lines of table here ending with \\
% \end{tabular}
% \end{ruledtabular}
% \end{table}

% Surround table environment with turnpage environment for landscape
% table
% \begin{table}
% \begin{ruledtabular}
% \end{tabular}
% \end{ruledtabular}
% \end{table}

%\appendix
%\section{}

%\begin{acknowledgments}
%\end{acknowledgments}


\bibliography{manusc.bib}

\end{document}